\documentclass[conference]{IEEEtran}

\usepackage{cite}
\usepackage{amsmath,amssymb,amsfonts}
\usepackage{graphicx}
\usepackage{textcomp}
\usepackage{xcolor}
\usepackage{booktabs}
\usepackage{multirow}
\usepackage{hyperref}
\usepackage{url}
\usepackage{microtype}

\hypersetup{
  colorlinks=true,
  linkcolor=black,
  citecolor=black,
  urlcolor=blue
}

\def\BibTeX{{\rm B\kern-.05em{\sc i\kern-.025em b}\kern-.08em
    T\kern-.1667em\lower.7ex\hbox{E}\kern-.125emX}}

\begin{document}

\title{How Many Tries Does It Take? Iterative Self-Repair in LLM Code Generation Across Model Scales and Benchmarks}

\author{\IEEEauthorblockN{Johin Johny Arimbur}
\IEEEauthorblockA{Independent Researcher \\
johinjohny144@gmail.com}}

\maketitle

\begin{abstract}
Large language models frequently fail to produce correct code on their first attempt, yet most benchmarks evaluate them in a single-shot setting. We investigate \emph{iterative self-repair} (feeding execution errors back to the model for correction) across seven models spanning three families and both open-weight and proprietary providers: Llama~3.1~8B, Llama~3.3~70B, Llama~4~Scout (MoE, 16 experts), Llama~4~Maverick (MoE, 128 experts), Qwen3~32B, Gemini~2.5~Flash, and Gemini~2.5~Pro. On HumanEval (164 problems) and MBPP Sanitized (257 problems) with up to five attempts, self-repair universally improves pass rates: +4.9 to +17.1\,pp on HumanEval and +16.0 to +30.0\,pp on MBPP. Gemini~2.5~Flash achieves the highest final pass rates (96.3\% HumanEval, 93.8\% MBPP). Most gains concentrate in the first two rounds. Error-type analysis shows assertion errors (logical mistakes) are the hardest to repair at $\sim$45\%, while syntax and name errors are repaired at substantially higher rates, connecting to broader findings on the limits of LLM self-correction. Prior work found that weaker models fail at self-repair or require fine-tuning; we show that modern instruction-tuned models succeed with prompting alone, even at 8B scale. We also provide the first comparison of dense and MoE architectures for self-repair, and extend the repair-vs-resampling tradeoff analysis to modern models. A prompt ablation reveals chain-of-thought repair yields up to +5.5\,pp additional self-repair gain (measured as improvement in repair delta) over minimal prompting for capable models.
\end{abstract}


\section{Introduction}
\label{sec:introduction}

The ability of large language models (LLMs) to generate functionally correct code from natural-language specifications has improved dramatically in recent years~\cite{chen2021evaluating, li2022competition, roziere2024code}. Benchmarks such as HumanEval~\cite{chen2021evaluating} and MBPP~\cite{austin2021program} have become standard yardsticks, typically reporting a \emph{pass@1} metric: the fraction of problems solved correctly on a single attempt. While informative, this single-shot evaluation protocol diverges sharply from how human programmers actually write code. A professional developer routinely writes an initial draft, runs it, reads the error message, and iterates, often multiple times, before arriving at a correct solution.

\emph{Iterative self-repair} replicates this workflow within an LLM pipeline. When a generated program fails its test suite, the model receives the error traceback and is asked to produce a corrected version. This cycle can repeat for a fixed number of rounds. Prior work has explored self-repair with proprietary models such as GPT-3.5 and GPT-4~\cite{olausson2024selfrepair, chen2024teaching}, yielding mixed conclusions about its effectiveness. Olausson et al.~\cite{olausson2024selfrepair} notably argued that self-repair is ``not a silver bullet,'' observing that weaker models may introduce new errors during repair attempts.

However, the landscape has changed significantly since those studies. A new generation of open-weight models (2024--2025) exhibits substantially stronger instruction-following and error-comprehension capabilities, and novel architectures such as mixture-of-experts (MoE) have become prevalent. No prior work has evaluated self-repair on these modern models or compared dense vs.\ MoE architectures for repair. These developments motivate a fresh empirical investigation: \emph{does self-repair now work universally, even for smaller models? How do dense and MoE architectures differ in their repair dynamics? And what practical guidelines can we derive for deployment?}

In this paper, we address these questions with a systematic evaluation across seven models from three families: Llama~3.1~8B (8B dense)~\cite{grubisic2024llama3}, Llama~3.3~70B (70B dense)~\cite{llama33}, Llama~4~Scout~17B (17B active, MoE with 16 experts)~\cite{llama4}, Llama~4~Maverick~17B (17B active, MoE with 128 experts)~\cite{llama4}, Qwen3~32B (32B dense)~\cite{qwen3}, Gemini~2.5~Flash~\cite{gemini2025}, and Gemini~2.5~Pro~\cite{gemini2025}. The five open-weight models are accessed via the Groq free-tier API; the two Gemini models are accessed via Google Cloud Vertex AI. All use greedy decoding (temperature~$=0.0$) for reproducibility. We evaluate on two benchmarks, HumanEval (164 problems) and MBPP Sanitized (257 problems), with up to five attempts per problem.

Our contributions are as follows:
\begin{enumerate}
    \item We demonstrate that prompt-based self-repair is now effective across seven models from three families, even at 8B scale, updating the finding by Olausson et al.~\cite{olausson2024selfrepair} that weaker models are harmed by self-repair, and contrasting with Chen et al.~\cite{chen2025revisit} who found limited gains for a 7B model. Prior work required fine-tuning to enable small-model self-repair~\cite{ding2024cycle, jiang2024ledex}; we show that modern instruction-tuned models succeed without it.

    \item We present, to our knowledge, the first direct comparison of dense and mixture-of-experts architectures for self-repair across multiple model families, revealing distinct interaction patterns: on HumanEval, Scout (16 experts) achieves the highest repair gain among open-weight models (+14.0\,pp), while the Gemini models establish new ceiling performance levels.

    \item We provide cross-benchmark validation on both HumanEval and MBPP Sanitized, showing consistent model rankings and diminishing-returns patterns, strengthening generalizability beyond single-benchmark studies.

    \item We present fine-grained error-type analysis revealing that error category is a strong predictor of repair success: name errors are repaired at $\sim$77\% rates, while assertion errors succeed at only $\sim$45\%, connecting to the broader literature on LLM self-correction limits~\cite{huang2024large}.

    \item We quantify the diminishing-returns curve and derive practical deployment guidelines: two repair rounds capture the majority (76--95\%) of achievable gains. Our token-efficiency comparison against independent resampling extends the exploration-exploitation framework of Tang et al.~\cite{tang2024repair} to modern models, confirming that self-repair is increasingly advantageous for capable models.

    \item We conduct a repair prompt ablation comparing minimal, explain-then-fix, and chain-of-thought strategies, finding that CoT prompting yields +5.5\,pp additional repair gain over minimal prompting for the 70B model.
\end{enumerate}

\section{Related Work}
\label{sec:related}

\subsection{Code Generation with LLMs}

The emergence of Codex~\cite{chen2021evaluating} demonstrated that LLMs fine-tuned on code repositories could generate functionally correct programs from docstrings. Subsequent work scaled this capability through larger models~\cite{li2022competition}, code-specific pre-training~\cite{roziere2024code}, and reinforcement learning from execution feedback~\cite{le2022coderl}. The HumanEval benchmark~\cite{chen2021evaluating} and the Mostly Basic Programming Problems (MBPP) dataset~\cite{austin2021program} have become standard evaluation suites, while more challenging benchmarks such as APPS~\cite{hendrycks2021measuring} and SPoC~\cite{kulal2019spoc} test competition-level reasoning.

\subsection{Self-Debugging and Self-Repair}

Chen et al.~\cite{chen2024teaching} proposed \emph{Self-Debug}, demonstrating that LLMs can identify and fix bugs in their own code when provided with execution results. Their approach included several feedback strategies: simple error messages, code explanations, and unit test traces. Olausson et al.~\cite{olausson2024selfrepair} conducted a thorough investigation of self-repair, concluding that while GPT-4 benefited substantially, weaker models like GPT-3.5 often failed to improve or even degraded in performance. More recently, Chen et al.~\cite{chen2025revisit} revisited self-debugging with self-generated tests, finding that even a 7B model (Qwen2.5-Coder) showed limited or negative gains from prompt-based self-debugging on HumanEval. Two concurrent lines of work have shown that \emph{fine-tuning} can unlock self-repair for smaller models: CYCLE~\cite{ding2024cycle} trained 350M--3B models to self-refine through iterative code generation, and LeDex~\cite{jiang2024ledex} used supervised fine-tuning and reinforcement learning to enable self-debugging in CodeLlama-7B/13B. Our work differs from all of these in demonstrating that modern 2024--2025 instruction-tuned models benefit from prompt-based self-repair \emph{without any fine-tuning}, even at the 8B scale.

\subsection{Iterative Refinement Beyond Code}

The principle of iterative self-improvement has been explored in broader contexts. Madaan et al.~\cite{madaan2023selfrefine} introduced \emph{Self-Refine}, a general framework in which LLMs iteratively critique and revise their own outputs across diverse tasks. Shinn et al.~\cite{shinn2023reflexion} proposed \emph{Reflexion}, using verbal reinforcement signals to improve agent performance over multiple episodes. However, Huang et al.~\cite{huang2024large} cautioned that LLMs cannot reliably self-correct \emph{reasoning} without external feedback, an observation consistent with our finding that assertion errors (which require reasoning corrections) are the hardest to repair. Tang et al.~\cite{tang2024repair} formalized the repair-vs-resampling decision as an exploration-exploitation tradeoff, proposing a Thompson Sampling strategy (REx) that dynamically allocates between repair and resampling. Our resampling comparison extends this line of work with modern 2024--2025 models, confirming that the tradeoff is model-dependent.

\subsection{Multi-Turn and Collaborative Code Generation}

Recent work has also investigated multi-turn interactions for code generation. Key et al.~\cite{key2024multiturn} explored multi-turn code generation with instruction-tuned models, while Zhang et al.~\cite{zhang2023selfcollaboration} investigated self-collaboration patterns with ChatGPT. Zhong et al.~\cite{zhong2024ldb} proposed LDB, a debugger that verifies runtime execution step-by-step. SelfEvolve~\cite{jiang2023selfevolve} used LLMs to iteratively evolve code through self-generated feedback. Our work differs in its focus on the simplest possible self-repair protocol (feeding back raw error messages) and its systematic comparison across seven models from three families and two benchmarks.

\section{Methodology}
\label{sec:methodology}

\subsection{Self-Repair Protocol}

Our iterative self-repair protocol operates as follows. Given a programming problem specified by a function signature and docstring (as in HumanEval and MBPP), the model generates an initial code solution at round~$R_0$. This solution is extracted from the model's response, combined with the corresponding test cases, and executed in a sandboxed Python environment. If all test cases pass, the problem is marked as solved. If execution produces an error, we capture the error type and traceback message, then construct a repair prompt containing: (1)~the original problem specification, (2)~the model's previous code attempt, and (3)~the error message. The model is then asked to generate a corrected solution. This cycle repeats for up to four repair rounds ($R_1$ through~$R_4$), for a maximum of five total attempts per problem.

Formally, for a problem~$p$ and model~$M$, let $c_0 = M(p)$ be the initial code generation. If $\texttt{exec}(c_0)$ succeeds, we record success at~$R_0$. Otherwise, let $e_0 = \texttt{error}(c_0)$ be the captured error. For each subsequent round $i \in \{1, 2, 3, 4\}$:
\begin{equation}
    c_i = M(p, c_{i-1}, e_{i-1})
\end{equation}
where the model receives the problem, its previous attempt, and the error message. If $\texttt{exec}(c_i)$ succeeds at any round, the problem is marked as solved at round~$R_i$ and no further attempts are made.

\subsection{Code Extraction}

LLM outputs frequently contain natural-language explanations, markdown formatting, and other non-code content surrounding the actual solution. We employ a code extraction pipeline that: (1)~removes chain-of-thought traces (e.g., \verb|<think>...</think>| tags) that reasoning models may prepend; (2)~extracts code from markdown fences (\verb|```python ... ```|), including handling of unclosed fences from truncated responses; and (3)~if no complete function definition is found, prepends the function signature to body-only responses with appropriate indentation.

\subsection{Error Classification}

When a solution fails, we classify the error into one of the following categories based on the Python exception type:
\begin{itemize}
    \item \textbf{AssertionError}: The code runs without crashing but produces incorrect output, failing one or more test assertions. These errors indicate logical mistakes.
    \item \textbf{SyntaxError}: The generated code is not valid Python.
    \item \textbf{TypeError / ValueError}: Runtime type mismatches or invalid values, often indicating misunderstanding of the expected input/output types.
    \item \textbf{NameError}: References to undefined variables or functions, typically caused by incomplete code generation.
    \item \textbf{IndexError / KeyError}: Out-of-bounds access or missing dictionary keys, indicating edge-case handling failures.
    \item \textbf{Timeout}: The code does not terminate within the allotted time, suggesting infinite loops or excessive computational complexity.
\end{itemize}

\subsection{Metrics}

Our primary metric is \emph{cumulative pass@1} at each round~$R_i$, defined as the fraction of problems solved by round~$R_i$ (inclusive of all preceding rounds). Because each problem receives exactly one attempt per round with greedy decoding (temperature~$=0.0$), pass@1 is deterministic and does not require the unbiased estimator used for stochastic sampling~\cite{chen2021evaluating}. We also report the \emph{self-repair gain}~$\Delta$, defined as the difference between the final cumulative pass@1 (at~$R_4$) and the initial pass@1 (at~$R_0$), measured in percentage points~(pp). We note that on HumanEval ($n=164$), a single problem corresponds to $\approx$0.6\,pp; differences smaller than $\sim$2\,pp (3 problems) should be interpreted cautiously.

\section{Experimental Setup}
\label{sec:setup}

\subsection{Models}

We evaluate seven models from three families, spanning open-weight and proprietary providers:

\subsubsection{Open-Weight Models (Groq API)}

\begin{enumerate}
    \item \textbf{Llama 3.1 8B}~\cite{grubisic2024llama3}: An 8-billion-parameter dense transformer from Meta. This serves as our smallest baseline and represents the class of efficient, instruction-tuned models commonly deployed in resource-constrained settings.

    \item \textbf{Llama 3.3 70B}~\cite{llama33}: A 70-billion-parameter dense transformer from Meta. This is the largest dense open-weight model in our evaluation and provides a strong baseline for comparing against mixture-of-experts architectures.

    \item \textbf{Llama 4 Scout 17B}~\cite{llama4}: A mixture-of-experts (MoE) model from Meta with 17 billion active parameters distributed across 16 experts. The MoE architecture allows the model to maintain a large total parameter count while activating only a fraction of parameters per token.

    \item \textbf{Llama 4 Maverick 17B}~\cite{llama4}: A mixture-of-experts model from Meta with 17 billion active parameters and 128 experts. Compared to Scout, Maverick uses significantly more experts, providing an interesting comparison point for how MoE granularity affects code generation and self-repair.

    \item \textbf{Qwen3 32B}~\cite{qwen3}: A 32-billion-parameter dense transformer from Alibaba. Qwen3 supports both reasoning (thinking) and non-reasoning modes. Thinking mode was disabled via the \texttt{/no\_think} setting to ensure fair comparison with non-reasoning models.
\end{enumerate}

\subsubsection{Proprietary Models (Vertex AI)}

\begin{enumerate}
    \setcounter{enumi}{5}
    \item \textbf{Gemini 2.5 Flash}~\cite{gemini2025}: A lightweight model from Google optimized for speed and cost-efficiency. Gemini~2.5~Flash is designed as a fast-inference model suitable for high-throughput applications, making it an interesting comparison for self-repair efficiency.

    \item \textbf{Gemini 2.5 Pro}~\cite{gemini2025}: Google's flagship reasoning model with extended ``thinking'' capabilities. Unlike Qwen3's optional thinking mode, Gemini~2.5~Pro integrates reasoning natively. This model represents the frontier of proprietary model capability in our evaluation.
\end{enumerate}

The inclusion of Gemini models extends our evaluation from two model families (Meta, Alibaba) to three (Meta, Alibaba, Google), and from open-weight-only to a mix of open-weight and proprietary models. This broadens the generalizability of our findings.

All models are queried with greedy decoding (temperature~$=0.0$) to ensure deterministic, reproducible outputs. A minimal system prompt is used for all models: \emph{``You are an expert Python programmer. Complete the given function. Return ONLY the Python code, no explanations, no markdown formatting.''} No few-shot examples are provided. For initial generation, each model receives only the function signature and docstring. For repair rounds, the model receives the original problem, its previous code attempt, and the error message. The repair prompt is minimal: it presents the error and asks for a corrected function (see Section~\ref{sec:methodology}).

\subsection{Benchmarks}

We evaluate on two standard code generation benchmarks:

\begin{enumerate}
    \item \textbf{HumanEval}~\cite{chen2021evaluating}: 164 hand-written Python programming problems, each with a function signature, docstring, and test cases. Problems range from simple string manipulations to moderately complex algorithmic tasks. All 164 problems were evaluated for all seven models.

    \item \textbf{MBPP Sanitized}~\cite{austin2021program}: 257 problems from the Mostly Basic Programming Problems dataset (sanitized subset). MBPP problems tend to be shorter and more focused than HumanEval, but cover a broader range of programming concepts. All 257 problems were evaluated for all seven models.
\end{enumerate}

We acknowledge that these benchmarks are considered relatively easy by 2025 standards. We discuss the implications of this choice and the need for evaluation on harder benchmarks in Section~\ref{sec:limitations}.

\subsection{Infrastructure}

Open-weight model experiments were conducted using the Groq free-tier API, which provides rate-limited access to hosted models at no monetary cost. Gemini experiments were conducted using Google Cloud Vertex AI, with total Gemini API costs under \$20 USD (including initial debugging runs). Code execution was performed in isolated Python subprocesses with a 15-second timeout per test case. All code and results are publicly available.\footnote{\url{https://github.com/Johin2/iterative-code-repair}}

\section{Results}
\label{sec:results}

\subsection{HumanEval Results}

\begin{table}[t]
\centering
\caption{Cumulative pass@1 (\%) on HumanEval (164 problems) by repair round. $R_0$ is the initial attempt; $R_1$--$R_4$ are repair rounds. $\Delta$ denotes the total self-repair gain ($R_4 - R_0$) in percentage points.}
\label{tab:humaneval}
\resizebox{\columnwidth}{!}{%
\begin{tabular}{llcccccc}
\toprule
\textbf{Model} & \textbf{Family} & $R_0$ & $R_1$ & $R_2$ & $R_3$ & $R_4$ & $\Delta$ \\
\midrule
Llama 3.1 8B     & Meta & 67.1 & 73.2 & 75.6 & 76.2 & 76.8 & +9.8 \\
Llama 3.3 70B    & Meta & 82.9 & 89.6 & 90.9 & 92.7 & 93.3 & +10.4 \\
Scout 17B (16E)  & Meta & 75.6 & 84.8 & 87.2 & 89.0 & 89.6 & +14.0 \\
Maverick 17B (128E) & Meta & 87.2 & 91.5 & 92.7 & 93.9 & 93.9 & +6.7 \\
Qwen3 32B        & Alibaba & 87.8 & 90.2 & 92.1 & 92.7 & 92.7 & +4.9 \\
\midrule
Gemini 2.5 Flash & Google & 86.6 & 95.7 & 95.7 & 96.3 & 96.3 & +9.8 \\
Gemini 2.5 Pro   & Google & 73.2 & 82.9 & 86.6 & 89.6 & 90.2 & +17.1 \\
\bottomrule
\end{tabular}%
}
\end{table}

Table~\ref{tab:humaneval} presents the cumulative pass@1 rates on HumanEval across all repair rounds for all seven models. Several key observations emerge:

\textbf{Universal improvement.} Self-repair improves pass rates for every model tested, across all three families and both open-weight and proprietary models. Every model gains at least +4.9\,pp.

\textbf{Gemini~2.5~Flash achieves the highest final rate.} At 96.3\%, Flash surpasses all open-weight models, including Maverick (93.9\%) and Llama~3.3~70B (93.3\%). Its R1 jump from 86.6\% to 95.7\% (+9.1\,pp in a single round) is among the largest first-round gains (comparable to Pro's +9.8\,pp and matching Scout's +9.1\,pp), suggesting strong error comprehension.

\textbf{Gemini~2.5~Pro shows the largest repair gain.} Pro achieves +17.1\,pp, the highest $\Delta$ on HumanEval. However, its lower initial pass rate (73.2\%) reflects overhead from its native reasoning traces, which embed extended chain-of-thought within the response and occasionally interfere with code extraction despite our extraction pipeline (see Section~\ref{sec:methodology}). A similar pattern appears with Qwen3's thinking mode (see Section~\ref{sec:limitations}). Pro's final rate (90.2\%) is competitive but below Flash, illustrating that stronger reasoning does not always translate to better code generation in a simple prompt-completion paradigm.

\textbf{Diminishing returns from higher baselines.} Among the open-weight models, the pattern holds: Qwen3~32B starts at 87.8\% and gains only +4.9\,pp, while Scout starts lower at 75.6\% and gains +14.0\,pp. Flash is a notable exception: it starts high (86.6\%) yet still achieves a substantial +9.8\,pp gain, suggesting that its repair capability exceeds the diminishing-returns trend.

\begin{figure}[t]
    \centering
    \includegraphics[width=\columnwidth]{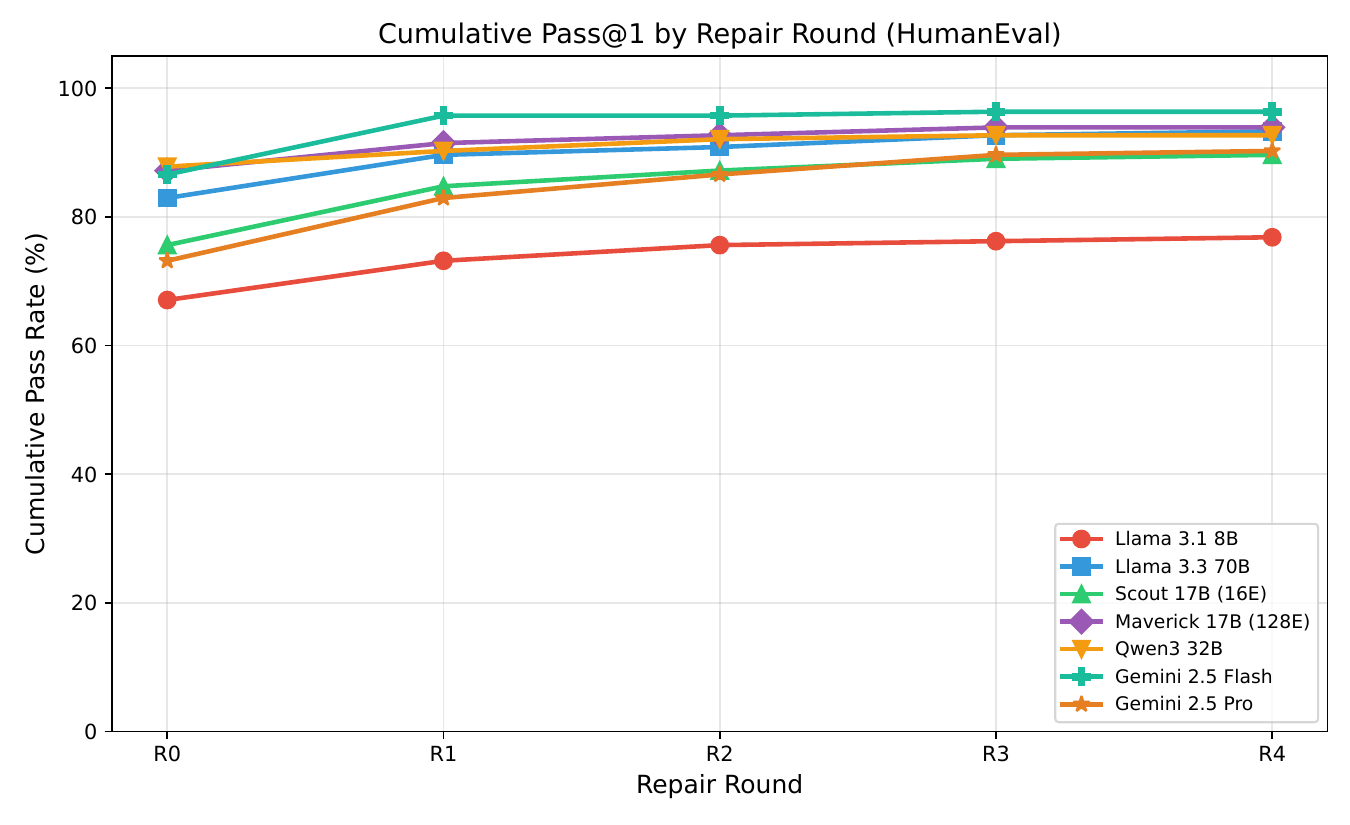}
    \caption{Cumulative pass@1 rate on HumanEval across repair rounds for all seven models. Gemini~2.5~Flash achieves the highest final rate (96.3\%), while most gains concentrate in the first two rounds.}
    \label{fig:cumulative}
\end{figure}

Figure~\ref{fig:cumulative} visualizes these trajectories. The curves illustrate the universal upward trend, with Flash breaking through the $\sim$94\% ceiling observed among the open-weight models.

\subsection{MBPP Results}

\begin{table}[t]
\centering
\caption{Cumulative pass@1 (\%) on MBPP Sanitized (257 problems) by repair round.}
\label{tab:mbpp}
\resizebox{\columnwidth}{!}{%
\begin{tabular}{llcccccc}
\toprule
\textbf{Model} & \textbf{Family} & $R_0$ & $R_1$ & $R_2$ & $R_3$ & $R_4$ & $\Delta$ \\
\midrule
Llama 3.1 8B     & Meta & 55.6 & 66.9 & 69.6 & 70.8 & 71.6 & +16.0 \\
Llama 3.3 70B    & Meta & 67.7 & 86.4 & 89.5 & 89.9 & 90.7 & +23.0 \\
Scout 17B (16E)  & Meta & 65.4 & 77.4 & 80.5 & 82.9 & 83.3 & +17.9 \\
Maverick 17B (128E) & Meta & 72.0 & 85.6 & 88.7 & 91.8 & 92.6 & +20.6 \\
Qwen3 32B        & Alibaba & 70.8 & 81.7 & 85.6 & 87.9 & 88.3 & +17.5 \\
\midrule
Gemini 2.5 Flash & Google & 63.8 & 86.8 & 91.8 & 93.4 & 93.8 & +30.0 \\
Gemini 2.5 Pro   & Google & 66.5 & 85.2 & 89.1 & 92.2 & 92.2 & +25.7 \\
\bottomrule
\end{tabular}%
}
\end{table}

Table~\ref{tab:mbpp} presents results on MBPP Sanitized. The MBPP results reinforce and extend the findings from HumanEval:

\textbf{Larger absolute gains.} Self-repair gains on MBPP are substantially larger than on HumanEval for every model. Among open-weight models, gains range from +16.0\,pp to +23.0\,pp. The Gemini models show even larger gains: Flash achieves +30.0\,pp and Pro +25.7\,pp, the two highest in our study.

\textbf{Flash dominates on MBPP.} Gemini~2.5~Flash achieves the highest final pass rate (93.8\%), despite starting at only 63.8\%, lower than four of the five open-weight models. Its +30.0\,pp repair gain exceeds the best open-weight model on MBPP (+23.0\,pp) by 7\,pp, suggesting qualitatively different repair behavior. Flash's R1 jump (+23.0\,pp in a single round) accounts for the majority of its gain.

\textbf{Consistent model ranking.} The relative ordering of models is largely consistent across benchmarks. Llama~3.1~8B has the lowest final rate on both benchmarks, while Flash leads on both. On MBPP ($n=257$), one problem corresponds to $\sim$0.4\,pp, providing finer resolution for ranking comparisons.

\subsection{Cross-Benchmark Comparison}

\begin{table}[t]
\centering
\caption{Cross-benchmark comparison of self-repair gains ($\Delta$, in percentage points) on HumanEval and MBPP Sanitized.}
\label{tab:cross}
\begin{tabular}{llcc}
\toprule
\textbf{Model} & \textbf{Family} & \textbf{HumanEval $\Delta$} & \textbf{MBPP $\Delta$} \\
\midrule
Llama 3.1 8B     & Meta & +9.8  & +16.0 \\
Llama 3.3 70B    & Meta & +10.4 & +23.0 \\
Scout 17B (16E)  & Meta & +14.0 & +17.9 \\
Maverick 17B (128E) & Meta & +6.7  & +20.6 \\
Qwen3 32B        & Alibaba & +4.9  & +17.5 \\
\midrule
Gemini 2.5 Flash & Google & +9.8  & +30.0 \\
Gemini 2.5 Pro   & Google & +17.1 & +25.7 \\
\bottomrule
\end{tabular}

\end{table}

Table~\ref{tab:cross} summarizes the self-repair gains across both benchmarks. MBPP consistently yields higher repair gains than HumanEval for every model. This difference likely reflects two factors: (1) lower initial pass rates on MBPP provide more room for improvement, and (2) MBPP problems, while numerous, tend to be shorter and more focused, making their errors potentially easier to diagnose and repair. The Gemini models show the largest MBPP gains (+30.0 and +25.7\,pp), suggesting that their error comprehension scales particularly well on shorter, focused problems.

\begin{figure}[t]
    \centering
    \includegraphics[width=\columnwidth]{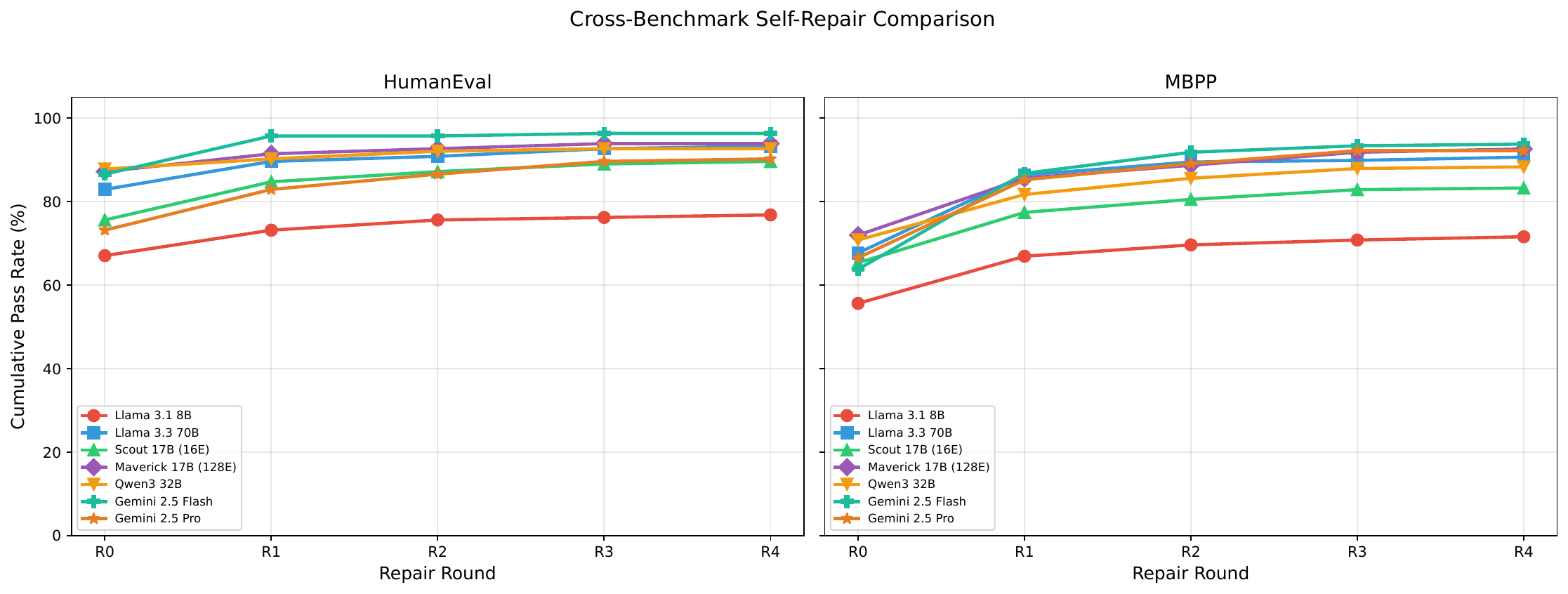}
    \caption{Cross-benchmark comparison of self-repair gains for all seven models. MBPP consistently yields larger gains than HumanEval, with Gemini~2.5~Flash showing the largest MBPP gain (+30.0\,pp).}
    \label{fig:cross}
\end{figure}

Figure~\ref{fig:cross} visualizes the cross-benchmark comparison, highlighting the consistent pattern of larger MBPP gains across all models and families.

\subsection{Error Distribution Analysis}

\begin{figure}[t]
    \centering
    \includegraphics[width=\columnwidth]{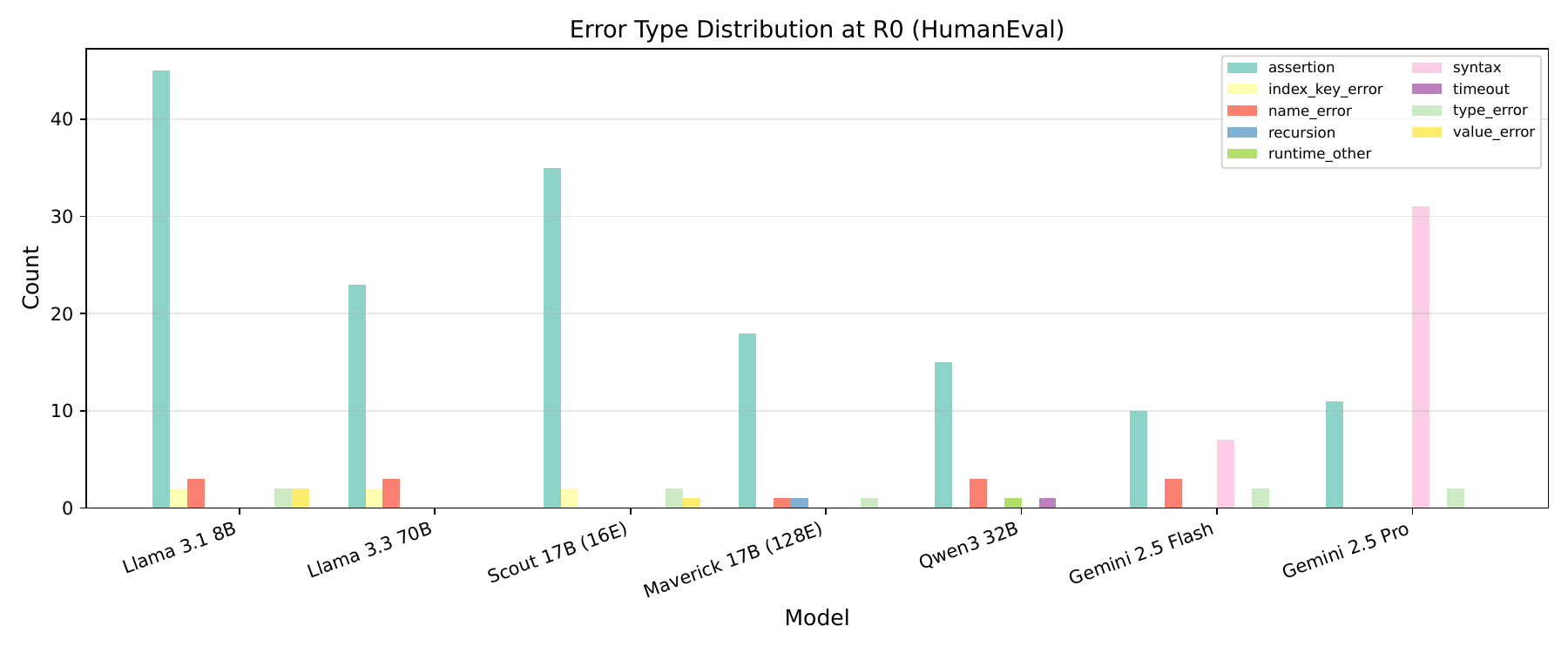}
    \caption{Distribution of error types at the initial attempt ($R_0$) on HumanEval for each model.}
    \label{fig:errors}
\end{figure}

Figure~\ref{fig:errors} shows the distribution of error types at the initial attempt ($R_0$) on HumanEval. Across all seven models, assertion errors dominate, indicating that most failures produce syntactically valid but logically incorrect code. Higher-capability models have fewer total failures but the same error-type profile, confirming that assertion errors are the irreducible core of failure regardless of model scale or family.

\subsection{Repair Success by Error Type}

\begin{figure}[t]
    \centering
    \includegraphics[width=\columnwidth]{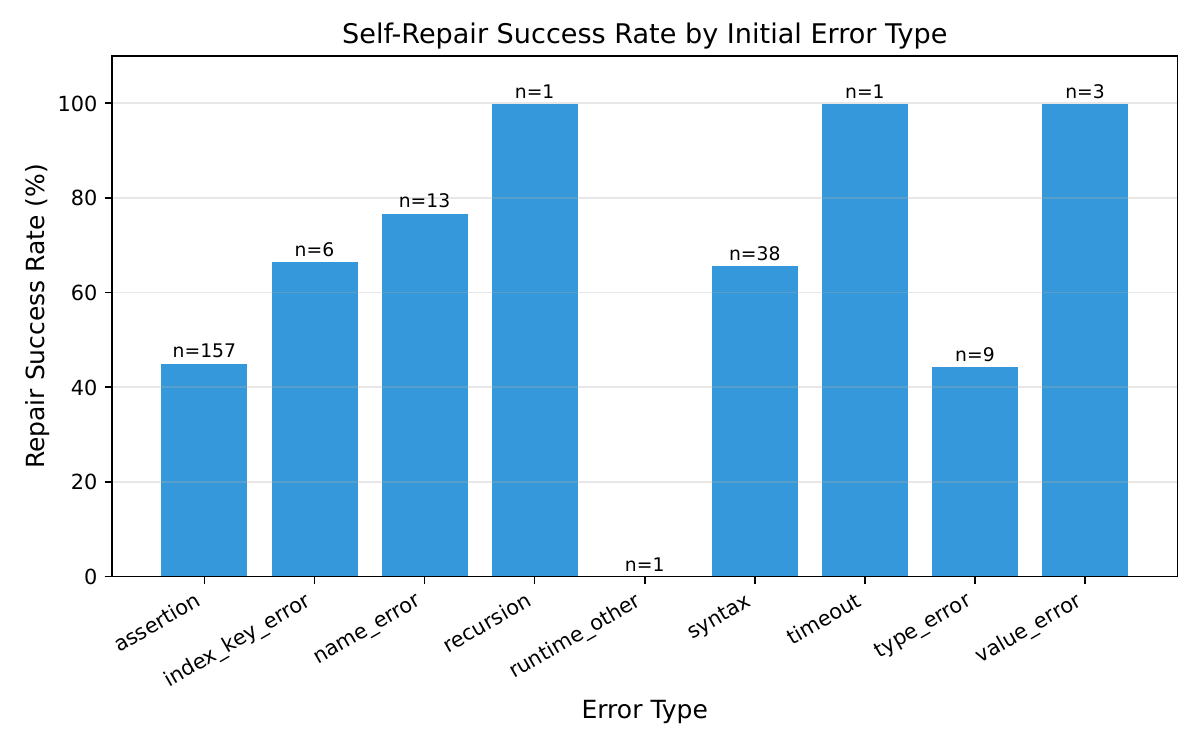}
    \caption{Repair success rate by error type across all models. Assertion errors (logical mistakes) are the hardest to repair.}
    \label{fig:repair_by_error}
\end{figure}

Figure~\ref{fig:repair_by_error} analyzes repair success rates broken down by error type. A clear hierarchy emerges:

\begin{itemize}
    \item \textbf{Name errors} are repaired at high rates ($\sim$77\%). These errors provide specific diagnostic information (undefined variable or function names) that guides the model toward the fix.

    \item \textbf{Syntax errors} are repaired at moderate rates ($\sim$66\%). While many syntax errors are straightforward to fix, some reflect deeper structural issues in code generation (e.g., incomplete functions or mismatched indentation) that are harder to resolve.

    \item \textbf{Assertion errors} are the hardest to repair, with success rates around 45\%. An assertion error indicates that the code ran successfully but produced the wrong output. The error message provides minimal diagnostic information about \emph{what} the code computed incorrectly or \emph{why}, requiring the model to re-examine its logic.
\end{itemize}

This finding is consistent with Huang et al.'s observation~\cite{huang2024large} that LLMs struggle to self-correct reasoning without external feedback. On MBPP, assertion errors are repaired at higher rates ($\sim$63\%) than on HumanEval ($\sim$45\%), likely because MBPP's shorter problems make logical errors easier to diagnose.

\subsection{Per-Round Improvement}

\begin{figure}[t]
    \centering
    \includegraphics[width=\columnwidth]{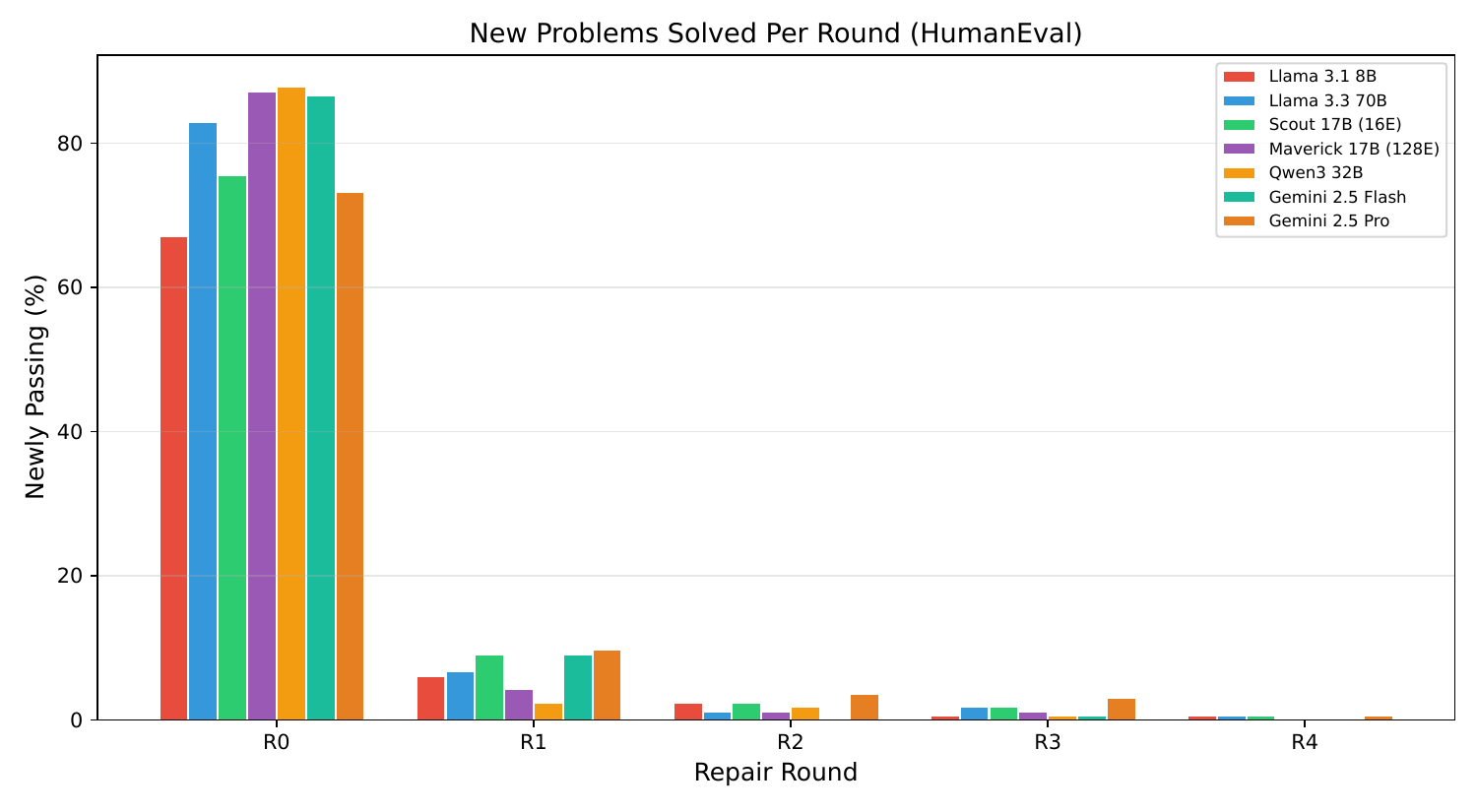}
    \caption{Marginal improvement in pass@1 per repair round on HumanEval. The first repair round ($R_0 \to R_1$) yields the largest gain for all models.}
    \label{fig:per_round}
\end{figure}

Figure~\ref{fig:per_round} shows the marginal improvement at each repair round on HumanEval. The pattern of diminishing returns is consistent across all models:

\begin{itemize}
    \item \textbf{$R_0 \to R_1$ (first repair)}: The largest single-round gain for all seven models. Gemini~2.5~Pro gains +9.8\,pp (the largest first-round gain), followed by Scout and Flash (both +9.1\,pp), while Qwen3 gains +2.4\,pp (the smallest, reflecting its high baseline).

    \item \textbf{$R_1 \to R_2$ (second repair)}: Still meaningful but reduced for all models.

    \item \textbf{$R_2 \to R_3$ and $R_3 \to R_4$}: Minimal additional gains. For Qwen3, Maverick, and Gemini~2.5~Flash, $R_4$ yields no additional improvement over~$R_3$, indicating that these models have solved all problems amenable to self-repair by round~3.
\end{itemize}

The concentration of gains in early rounds has practical implications: for most applications, two repair rounds capture the bulk of the benefit, and additional rounds offer diminishing marginal returns relative to their computational cost.

\subsection{Token Cost Analysis}

\begin{table}[t]
\centering
\caption{Total token usage (prompt $+$ completion) and self-repair gain on HumanEval (164 problems). Gemini models' internal reasoning tokens are excluded for comparability; see text.}
\label{tab:tokens}
\begin{tabular}{lccc}
\toprule
\textbf{Model} & \textbf{Total Tokens} & $\Delta$ \textbf{(pp)} & \textbf{Tokens/pp} \\
\midrule
Llama 3.1 8B     & 195K & +9.8  & 19.9K \\
Llama 3.3 70B    & 112K & +10.4 & 10.8K \\
Scout 17B (16E)  & 155K & +14.0 & 11.1K \\
Maverick 17B (128E) & 93K  & +6.7  & 13.9K \\
Qwen3 32B        & 102K & +4.9  & 20.8K \\
\midrule
Gemini 2.5 Flash & 121K & +9.8  & 12.3K \\
Gemini 2.5 Pro   & 162K & +17.1 & 9.5K \\
\bottomrule
\end{tabular}
\end{table}

Table~\ref{tab:tokens} reports total token usage for each model on HumanEval. Token counts reflect prompt and completion tokens only. For Gemini models, internal reasoning (``thinking'') tokens are excluded from this count; including them would increase Flash's total to $\sim$489K and Pro's to $\sim$994K. Since thinking tokens are not directly comparable to standard input/output tokens, we report prompt+completion for fair cross-model comparison. Gemini~2.5~Pro achieves the best cost-effectiveness at 9.5K tokens per percentage point of improvement, followed by Llama~3.3~70B at 10.8K tokens/pp.

\begin{table}[t]
\centering
\caption{Per-round repair breakdown on HumanEval: number of problems first solved at each round, and the repair success rate (fraction of initially-failed problems eventually repaired).}
\label{tab:perround}
\resizebox{\columnwidth}{!}{%
\begin{tabular}{lccccccc}
\toprule
\textbf{Model} & $R_0$ & $R_1$ & $R_2$ & $R_3$ & $R_4$ & \textbf{Never} & \textbf{Rep.\,\%} \\
\midrule
Llama 3.1 8B     & 110 & 10 & 4 & 1 & 1 & 38 & 29.6 \\
Llama 3.3 70B    & 136 & 11 & 2 & 3 & 1 & 11 & 60.7 \\
Scout 17B (16E)  & 124 & 15 & 4 & 3 & 1 & 17 & 57.5 \\
Maverick 17B (128E) & 143 &  7 & 2 & 2 & 0 & 10 & 52.4 \\
Qwen3 32B        & 144 &  4 & 3 & 1 & 0 & 12 & 40.0 \\
\midrule
Gemini 2.5 Flash & 142 & 15 & 0 & 1 & 0 &  6 & 72.7 \\
Gemini 2.5 Pro   & 120 & 16 & 6 & 5 & 1 & 16 & 63.6 \\
\bottomrule
\end{tabular}%
}
\end{table}

Table~\ref{tab:perround} provides a fine-grained breakdown of when problems are first solved. The first repair round ($R_1$) accounts for the majority of all repairs across every model. Repair success rate (the fraction of initially-failed problems that are eventually fixed) correlates positively with model capability: Gemini~2.5~Flash repairs 72.7\% of its failures (the highest), followed by Gemini~2.5~Pro at 63.6\% and Llama~3.3~70B at 60.7\%, while Llama~3.1~8B repairs only 29.6\%.

\subsection{Repair Prompt Ablation}

Our main experiments use a minimal repair prompt that provides only the error message and asks for a corrected function. To assess sensitivity to prompt design, we compare up to three repair prompt strategies on four models: Llama~3.1~8B (dense, 8B), Llama~3.3~70B (dense, 70B), Llama~4~Scout (MoE, 17B active), and Qwen3~32B (dense, 32B). We evaluate all 164 HumanEval problems with up to two repair rounds ($R_0$--$R_2$). All three strategies were evaluated for three models; for Scout, only CoT and Explain-then-fix are available because the Minimal run was interrupted by Groq API rate limits and could not be re-run before the evaluation window closed.

\begin{enumerate}
    \item \textbf{Minimal}: The baseline strategy used throughout this paper. The repair prompt presents the error message and requests a corrected function.
    \item \textbf{Explain-then-fix}: The model is asked to first explain the bug in 1--2 sentences, then provide the corrected code. This follows the explain-before-fix paradigm from Self-Debug~\cite{chen2024teaching}.
    \item \textbf{Chain-of-thought (CoT)}: The model is prompted to reason step-by-step: (1) what does the error tell us, (2) what is the root cause, (3) what is the fix. The corrected code follows the analysis.
\end{enumerate}

\begin{table}[t]
\centering
\caption{Repair prompt ablation on HumanEval. Cumulative pass@1 (\%) at each round for up to three prompt strategies on four models. Scout's Minimal run is unavailable (see text). Deltas computed from raw problem counts before rounding. $R_0$ and subsequent rounds may vary slightly across runs due to API-level non-determinism; $\Delta$ computed from each run's own $R_0$.}
\label{tab:ablation}
\begin{tabular}{llcccc}
\toprule
\textbf{Model} & \textbf{Strategy} & $R_0$ & $R_1$ & $R_2$ & $\Delta$ \\
\midrule
Llama 3.1 8B & Chain-of-thought & 65.9 & 72.6 & \textbf{77.4} & +11.6 \\
             & Explain-then-fix & 65.9 & 72.0 & 76.2 & +10.4 \\
             & Minimal          & 65.9 & 73.2 & 75.6 & +9.8 \\
\midrule
Llama 3.3 70B & Chain-of-thought & 84.1 & 93.9 & \textbf{96.3} & \textbf{+12.2} \\
              & Explain-then-fix & 82.9 & 90.2 & 93.9 & +11.0 \\
              & Minimal          & 81.1 & 86.6 & 87.8 & +6.7 \\
\midrule
Scout 17B (16E) & Chain-of-thought & 79.3 & 87.2 & \textbf{89.6} & +10.4 \\
                & Explain-then-fix & 78.0 & 87.2 & 87.2 & +9.1 \\
\midrule
Qwen3 32B & Chain-of-thought & 89.6 & 97.0 & \textbf{97.0} & +7.3 \\
          & Explain-then-fix & 88.4 & 93.3 & 94.5 & +6.1 \\
          & Minimal          & 87.8 & 90.2 & 91.5 & +3.7 \\
\bottomrule
\end{tabular}
\end{table}

\begin{figure}[t]
    \centering
    \includegraphics[width=\columnwidth]{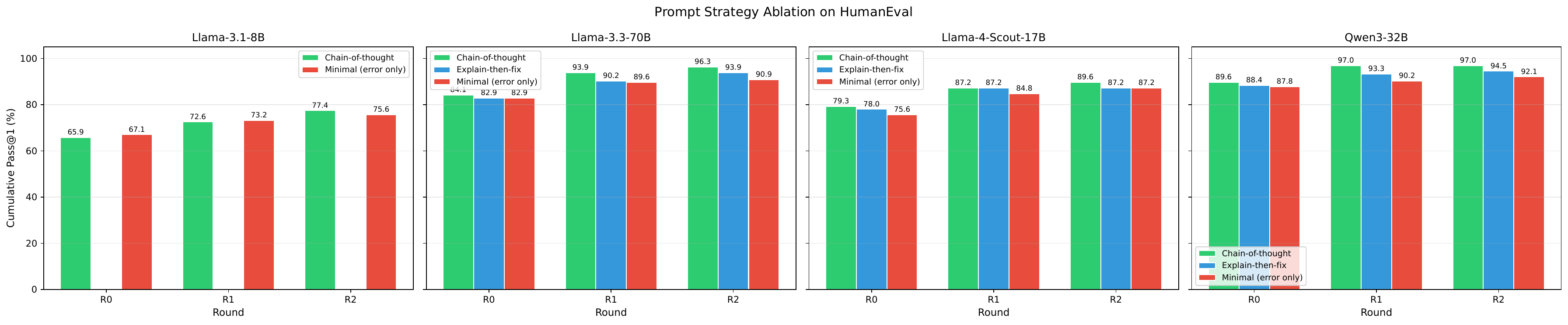}
    \caption{Prompt strategy ablation on HumanEval. CoT consistently achieves the highest final pass rate across all models, with the benefit scaling with model capability.}
    \label{fig:ablation}
\end{figure}

Table~\ref{tab:ablation} and Figure~\ref{fig:ablation} present the results across four models.

Note that each ablation configuration involves a complete re-execution of all 164 problems, including initial generation at $R_0$. Since the repair strategy does not affect the initial generation prompt, any $R_0$ variation across strategies for the same model reflects non-determinism in the Groq API's served model weights across sessions. For example, Llama~3.3~70B shows $R_0$ values of 84.1\%, 82.9\%, and 81.1\% across the three strategies. The $\Delta$ column is computed relative to each run's own $R_0$, making repair gain measurements independent of this baseline variation. However, run-to-run variation is non-trivial (e.g., 70B's $R_2$ ranges from 87.8\% to 96.3\%), and some of this gap may reflect session-level variance. Crucially, the key finding holds regardless of which $R_0$ baseline is used: CoT achieves the highest final $R_2$ rate for every model.

\textbf{CoT consistently wins across all models.} Chain-of-thought achieves the highest final pass rate for every model: 77.4\% (8B), 96.3\% (70B), 89.6\% (Scout), and 97.0\% (Qwen3). Qwen3 with CoT achieves the highest pass rate of any model-strategy combination in our ablation (97.0\% at $R_2$).

\textbf{The benefit of richer prompts scales with model capability.} Measured by the additional self-repair gain ($\Delta_\text{CoT} - \Delta_\text{minimal}$), the advantage of CoT is modest for the 8B model (+1.8\,pp) but widens for stronger models: +5.5\,pp for Llama~3.3~70B and +3.6\,pp for Qwen3. In absolute terms, CoT achieves final $R_2$ rates 1.8--8.5\,pp higher than minimal, though part of this gap reflects $R_0$ variation across runs (see above). This confirms the interaction between model capability and prompt strategy across the three models with complete ablation data, spanning 8B to 70B parameters.

The practical implication is clear: for capable models ($\geq$32B), chain-of-thought repair prompting is strongly recommended and can yield gains comparable to an additional 1--2 repair rounds. For smaller models, the benefit is present but modest, and the simpler minimal prompt remains a competitive choice.

\subsection{Self-Repair vs.\ Independent Resampling}

A key question is whether the token budget spent on iterative repair could be better used by simply drawing multiple independent samples. Tang et al.~\cite{tang2024repair} formalized this as an exploration-exploitation tradeoff and proposed a Thompson Sampling strategy (REx) to dynamically allocate between repair and resampling. We extend this analysis to modern 2024--2025 models by comparing self-repair (greedy decoding, $T\!=\!0$, up to 5 sequential attempts with error feedback) against independent resampling ($T\!=\!0.8$, 5 independent samples) on HumanEval. This comparison involves two differences: error feedback (present in repair, absent in resampling) and decoding temperature ($T\!=\!0$ vs.\ $T\!=\!0.8$). A fully controlled study would also require stochastic repair ($T\!>\!0$ with feedback); we leave this to future work. Nevertheless, this comparison addresses the practical question: given a token budget, should one invest in sequential repair or parallel resampling? For resampling, we report the unbiased pass@$k$ estimator~\cite{chen2021evaluating}.

\begin{table}[t]
\centering
\caption{Self-repair (greedy, 5 rounds) vs.\ independent resampling (temperature 0.8, 5 samples) on HumanEval (164 problems). Bold indicates higher pass rate. Note: repair uses greedy decoding ($T\!=\!0$) while resampling uses $T\!=\!0.8$; see text.}
\label{tab:resampling}
\begin{tabular}{lcccc}
\toprule
\textbf{Model} & \multicolumn{2}{c}{\textbf{Self-Repair}} & \multicolumn{2}{c}{\textbf{Resampling}} \\
\cmidrule(lr){2-3} \cmidrule(lr){4-5}
 & Final\,\% & Tokens & pass@5\,\% & Tokens \\
\midrule
Llama 3.1 8B    & 76.8 & 195K & \textbf{79.9} & 219K \\
Llama 3.3 70B   & \textbf{93.3} & 112K & 90.9 & 231K \\
Scout 17B (16E) & \textbf{89.6} & 155K & 86.0 & 255K \\
Qwen3 32B       & 92.7 & 102K & 92.7 & 223K \\
\bottomrule
\end{tabular}
\end{table}

\begin{figure}[t]
    \centering
    \includegraphics[width=\columnwidth]{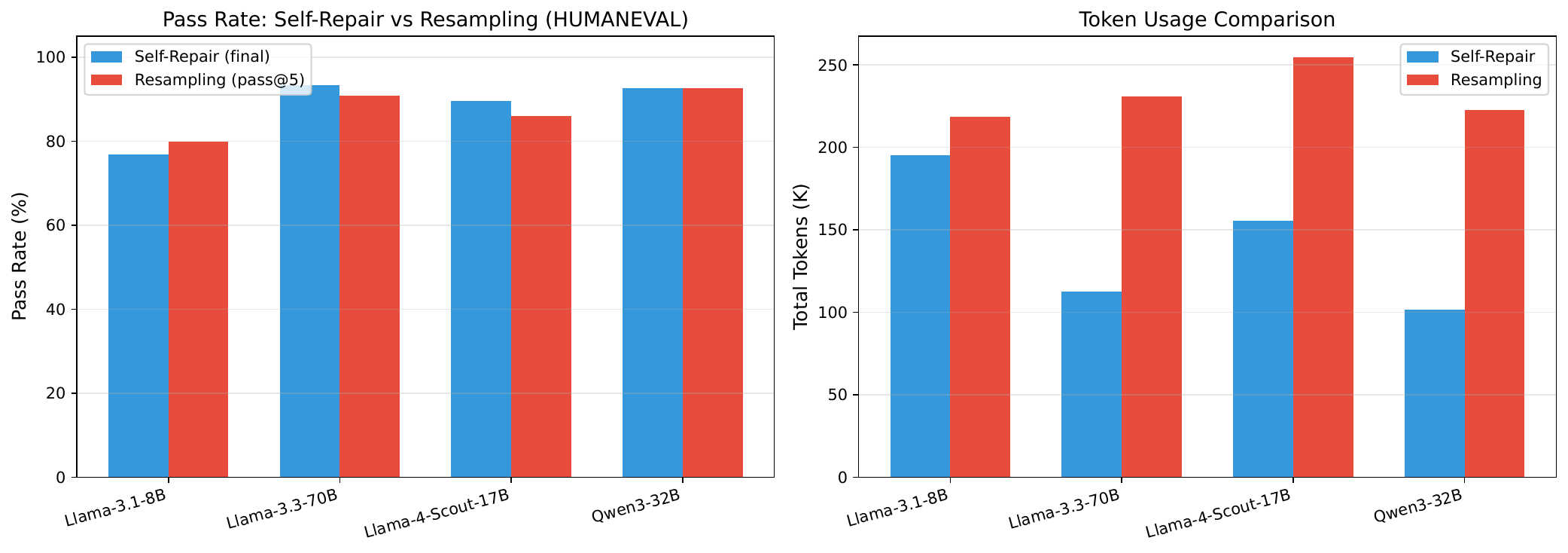}
    \caption{Self-repair vs.\ independent resampling on HumanEval. Self-repair matches or exceeds resampling for all models except the 8B, while consistently using fewer tokens.}
    \label{fig:resampling}
\end{figure}

Table~\ref{tab:resampling} and Figure~\ref{fig:resampling} compare self-repair against independent resampling on HumanEval for four models.\footnote{Maverick was unavailable on the Groq free-tier API during the resampling experiments; Gemini models were excluded because stochastic resampling on Vertex AI would have required additional paid API calls beyond our budget. The four evaluated models span the full capability range (8B--70B) and both architecture types (dense and MoE), providing sufficient evidence for the observed patterns.} The results reveal a clear, model-dependent pattern:

\textbf{Self-repair is always more token-efficient.} Across all four models, self-repair uses fewer tokens than resampling to achieve comparable or better pass rates, with savings ranging from 11\% (8B) to 54\% (Qwen3). The savings are largest for models whose strong error comprehension enables efficient single-attempt repairs, while the 8B model's longer repair traces narrow the gap. This efficiency advantage arises because self-repair generates only the corrected function at each round, while resampling generates a complete solution from scratch each time.

\textbf{Self-repair becomes increasingly competitive with model capability.} For the weakest model (Llama~3.1~8B), resampling achieves a modestly higher pass@5 (79.9\% vs.\ 76.8\%); the diversity of independent samples compensates for the model's limited error comprehension. For the three stronger models, self-repair matches or exceeds resampling: Scout (89.6\% vs.\ 86.0\%), Llama~3.3~70B (93.3\% vs.\ 90.9\%, a difference of 4 problems), and Qwen3~32B (92.7\% tied) all favor repair. The 70B result is particularly striking: self-repair achieves +2.4\,pp higher pass rate while using less than half the tokens (112K vs.\ 231K).

This pattern has a clear interpretation: as models become more capable, their ability to diagnose and fix errors from feedback improves, making the \emph{informational signal} from error messages more valuable than the \emph{diversity} from independent samples. We note that this comparison does not fully disentangle the value of error feedback from the effect of decoding temperature; a stochastic repair condition would be needed to isolate these factors. The practical implication is that self-repair is the preferred strategy for capable models, while a hybrid approach (repairing diagnosable errors and resampling for assertion errors) may be optimal for weaker models.

\section{Discussion}
\label{sec:discussion}

\subsection{Self-Repair as a Universal Improvement Strategy}

Our results demonstrate that iterative self-repair with error feedback is a universally effective strategy for improving LLM code generation across all seven models tested, spanning three families (Meta, Alibaba, Google), both open-weight and proprietary models, and both dense and MoE architectures. This finding updates the conclusions of Olausson et al.~\cite{olausson2024selfrepair}, who found that weaker models could be harmed by self-repair, and contrasts with Chen et al.~\cite{chen2025revisit}, who found limited gains for Qwen2.5-Coder-7B with prompt-based self-debugging. Prior work demonstrated that fine-tuning can enable small-model self-repair~\cite{ding2024cycle, jiang2024ledex}; our results suggest that the 2024--2025 generation of instruction-tuned models has closed this gap without task-specific training. The discrepancy with earlier findings likely reflects improvements in instruction-following and error-comprehension capabilities: all models in Olausson et al.'s study were released in 2022--2023, while ours cover 2024--2025.

The cross-benchmark consistency of our findings strengthens this conclusion. Self-repair improves every model on both HumanEval and MBPP, with largely consistent model rankings across benchmarks. The inclusion of Gemini models from a third family further strengthens generalizability: self-repair effectiveness is a robust property of modern LLMs rather than an artifact of a particular model family or benchmark.

\subsection{Dense vs.\ Mixture-of-Experts Architectures}

Our evaluation includes dense models (Llama 8B, 70B, Qwen3 32B) and MoE models (Scout 16E, Maverick 128E) among the open-weight models, as well as two proprietary Gemini models whose architectures have not been publicly disclosed, enabling comparisons across three families. Several observations are noteworthy:

\begin{itemize}
    \item Among open-weight models, Scout (16 experts) achieves the highest repair gain on HumanEval (+14.0\,pp), suggesting that its MoE architecture particularly benefits from the opportunity to repair its errors.

    \item Gemini~2.5~Flash achieves the highest final pass rates on both benchmarks (96.3\% HumanEval, 93.8\% MBPP), surpassing all open-weight models. Its combination of high initial accuracy and strong repair capability sets a new ceiling.

    \item Gemini~2.5~Pro, despite being a frontier reasoning model, underperforms Flash on both benchmarks (90.2\% vs.\ 96.3\% on HumanEval). This suggests that reasoning overhead can hurt in a simple prompt-completion paradigm, echoing our findings with Qwen3's thinking mode.

    \item The 70B dense model approaches Maverick's performance (93.3\% vs.\ 93.9\% on HumanEval; 90.7\% vs.\ 92.6\% on MBPP), suggesting that a large dense model can compete with a smaller MoE model.
\end{itemize}

These comparisons are suggestive but must be interpreted cautiously: models differ not only in architecture but also in training data, parameter counts, and other confounds that we cannot isolate. We report these observations as correlations rather than causal claims.

\subsection{Qualitative Analysis of Repair Outcomes}

To complement our quantitative findings, we examine representative cases from HumanEval.

\textbf{Successful repairs.} On HumanEval/38, Llama~3.3~70B produced code that failed at $R_0$ with a \texttt{NameError} and repaired it in a single round. More interestingly, on HumanEval/26, Llama~3.1~8B failed with assertion errors for three consecutive rounds before finally producing a correct solution at~$R_3$, demonstrating that persistence can resolve even logical errors. On HumanEval/77, Scout exhibited error-type mutation across rounds (type error $\rightarrow$ assertion $\rightarrow$ type error $\rightarrow$ pass), showing the model genuinely restructuring its approach rather than making superficial edits.

\textbf{Persistent failures.} On HumanEval/10, Llama~3.1~8B produced assertion errors on all five attempts, never resolving the underlying logic error. On HumanEval/32, Maverick exhibited error \emph{thrashing}: assertion $\rightarrow$ name error $\rightarrow$ assertion $\rightarrow$ assertion $\rightarrow$ timeout, the failure mode described by Olausson et al.~\cite{olausson2024selfrepair}. We identified several ``universally hard'' problems (e.g., HumanEval/132, HumanEval/145) that defeated all models even with four repair rounds.

\subsection{Preliminary Evaluation on Harder Benchmarks}

To assess whether self-repair generalizes beyond the relatively easy HumanEval and MBPP benchmarks, we conducted a preliminary evaluation on LiveCodeBench~\cite{jain2024livecodebench}, a contamination-free benchmark of competitive programming problems. We evaluated four models on 50 LiveCodeBench problems (18 easy, 23 medium, 9 hard) with the same 5-round self-repair protocol. The four models (8B, Scout, Qwen3, Flash) span the capability range; the remaining three were omitted due to API availability constraints. We emphasize that this evaluation is exploratory: the small sample size (50 problems) limits statistical power.

\begin{table}[t]
\centering
\caption{LiveCodeBench results (50 problems). Self-repair improves all four models evaluated.}
\label{tab:livecodebench}
\begin{tabular}{lccc}
\toprule
\textbf{Model} & $R_0$ & \textbf{Final} & $\Delta$ \\
\midrule
Llama 3.1 8B    & 10.0 & 12.0 & +2.0 \\
Scout 17B (16E) &  6.0 & 14.0 & +8.0 \\
Qwen3 32B       &  8.0 & 10.0 & +2.0 \\
Gemini 2.5 Flash &  0.0 & 16.0 & +16.0 \\
\bottomrule
\end{tabular}
\end{table}

Table~\ref{tab:livecodebench} shows that self-repair improves all four models on LiveCodeBench. Gemini~2.5~Flash achieves the largest gain (+16.0\,pp), reaching 16\% from an initial pass rate of 0\%; all solved problems come entirely from repair rounds. Flash's 0\% $R_0$ reflects the difficulty of competitive programming problems combined with stdin/stdout format requirements, yet its repair capability recovers 8 problems. Scout achieves the second-largest gain (+8.0\,pp), more than doubling its pass rate from 6\% to 14\%.

While the absolute pass rates are much lower than on HumanEval (as expected for competitive programming), self-repair remains effective. The relative repair gains are actually larger than on HumanEval, suggesting that self-repair \emph{may} scale to harder benchmarks where the lower baseline leaves more room for improvement. At $n=50$, each problem corresponds to 2\,pp, so differences smaller than $\sim$4\,pp should be interpreted cautiously.

\subsection{Practical Implications}

Our findings suggest several practical guidelines for deploying LLM code generation systems:

\begin{enumerate}
    \item \textbf{Always include at least one repair round}: The first repair round consistently provides the largest marginal improvement for all models tested.

    \item \textbf{Two rounds capture most gains}: For cost-sensitive applications, two repair rounds ($R_0$ through~$R_2$) capture the majority (76--95\%) of the total achievable improvement.

    \item \textbf{Consider error type for routing}: Name errors are repaired at the highest rates ($\sim$77\%), while assertion errors are the hardest ($\sim$45\%). For assertion-dominated failures, alternative strategies (e.g., sampling multiple independent solutions) may be more cost-effective.

    \item \textbf{Model selection matters}: For applications where final pass rate is paramount, Gemini~2.5~Flash (96.3\%) or Maverick (93.9\%) should be preferred. For self-repair gain per token, Gemini~2.5~Pro and Llama~3.3~70B offer the best efficiency.
\end{enumerate}

\section{Limitations and Future Work}
\label{sec:limitations}

We identify several limitations of this study that suggest directions for future work.

\textbf{Resampling comparison scope.}
Our resampling comparison (Section~\ref{sec:results}, Table~\ref{tab:resampling}) covers four of seven models on HumanEval only. While the results clearly show that self-repair is more token-efficient and increasingly advantageous for stronger models, extending this comparison to MBPP and controlling for token budget more precisely (e.g., allocating the exact same number of tokens to each strategy) would strengthen the analysis. Additionally, we compare greedy repair against stochastic resampling (temperature~$=0.8$), which conflates two differences: error feedback vs.\ diversity, and greedy vs.\ stochastic decoding. A fairer comparison might use stochastic repair as well.

\textbf{Benchmark scope.}
Our main evaluation uses HumanEval and MBPP, which are considered relatively easy by 2025 standards. Our preliminary LiveCodeBench evaluation (Section~\ref{sec:discussion}) shows that self-repair remains effective on harder competitive programming problems, but covers only four models on 50 problems. A comprehensive evaluation across all models on LiveCodeBench~\cite{jain2024livecodebench}, BigCodeBench~\cite{zhuo2024bigcodebench}, or SWE-Bench~\cite{jimenez2024swebench} would more fully characterize self-repair on problems requiring deep algorithmic reasoning.

\textbf{Greedy decoding only.}
Our use of temperature~$=0.0$ ensures deterministic, reproducible results but eliminates sampling variance, precluding confidence intervals or statistical significance tests. On HumanEval (164 problems), a single problem corresponds to $\pm$0.6\,pp, meaning differences of $\sim$1.2\,pp or less (e.g., Maverick at 93.9\% vs.\ Qwen3 at 92.7\%, a gap of just two problems) should not be over-interpreted. Experiments with stochastic decoding across multiple seeds would provide more robust estimates and enable statistical analysis.

\textbf{Ablation scope.}
Our prompt ablation covers four of seven models (8B, 70B, Scout, Qwen3) with two repair rounds ($R_0$--$R_2$). Extending to all five repair rounds would characterize whether CoT's advantage persists or converges in later rounds. The ablation uses HumanEval only; prompt sensitivity on MBPP or LiveCodeBench may differ.

\textbf{API-served model weights.}
Open-weight models were accessed via the Groq free-tier API, which may serve quantized or otherwise optimized model variants. We did not verify that the served weights exactly match the canonical model releases. Gemini models were accessed via Google Cloud Vertex AI, where the served model versions are controlled by Google. Pass rates may differ slightly across API versions or on locally-hosted weights.

\textbf{Model family diversity.} Our evaluation covers three model families (Llama, Qwen, and Gemini) across two inference providers (Groq and Vertex AI), including both open-weight and proprietary models. Code-specialized models (e.g., DeepSeek-Coder~\cite{guo2024deepseek}) may exhibit different repair dynamics due to code-focused pre-training and remain a direction for future work.

\textbf{Qwen3 reasoning mode.}
We evaluated Qwen3 with thinking mode enabled on all 164 HumanEval problems. Thinking mode achieves a \emph{lower} final pass rate (89.0\% vs.\ 92.7\% for no-think), despite a much larger repair gain (+14.0\,pp vs.\ +4.9\,pp). The explanation lies in the initial pass rate: thinking mode's $R_0$ drops to 75.0\% (vs.\ 87.8\% for no-think) because long reasoning traces contaminate the code output, causing syntax errors. Self-repair compensates by fixing 23 additional problems, but 18 problems fail due to API errors or persistent extraction issues. Our decision to disable thinking mode in the main experiments is validated: with the current code extraction pipeline, thinking mode reduces net performance. A more robust extraction approach (or native tool-use integration) could potentially unlock the reasoning benefits without the extraction overhead.

\section{Threats to Validity}
\label{sec:threats}

\textbf{Internal validity.}
Our use of greedy decoding ensures deterministic, reproducible results but reflects a single point in the sampling distribution. Prior work~\cite{olausson2024selfrepair, chen2024teaching} similarly reports single-run results under fixed decoding settings. Our prompt ablation (Table~\ref{tab:ablation}) shows that repair prompt design does affect outcomes, meaning the main results represent a lower bound for capable models. Additionally, minor $R_0$ variation in the ablation table suggests the Groq API may serve slightly different model versions across sessions.

\textbf{External validity.}
Our model selection (seven models from three families) spans a meaningful range, including both open-weight and proprietary models, but may not capture behaviors of all architectures. We do not evaluate code-specialized models (e.g., DeepSeek-Coder) or models with native tool use. Additionally, because HumanEval and MBPP are widely known, evaluated models may have been exposed to these problems during pre-training, which could inflate initial pass rates. However, this affects all models equally and does not invalidate relative comparisons.

\textbf{Construct validity.}
Our cumulative pass@1 metric counts a problem as solved if it passes all provided test cases. However, test suites may not cover all edge cases, meaning some ``passing'' solutions could contain latent bugs. Our error classification relies on Python exception types, which may conflate distinct failure modes.

\section{Conclusion}
\label{sec:conclusion}

We have presented a systematic study of iterative self-repair across seven models from three families (Meta, Alibaba, Google), spanning both open-weight and proprietary models, dense and MoE architectures, and two benchmarks (HumanEval and MBPP Sanitized). Self-repair is universally effective: every model improves on every benchmark. Error-type analysis confirms that name errors are repaired at high rates, while assertion errors remain the most challenging, connecting to broader findings on the limits of LLM self-correction.

Two repair rounds capture the majority of achievable gains, providing clear deployment guidance. Our prompt ablation shows that chain-of-thought repair can yield additional gains for capable models, though the benefit is model-dependent. These results establish iterative self-repair as a practical, training-free technique for improving LLM code generation, while identifying code-specialized models and harder-benchmark evaluation as important directions for future work.

\section*{Acknowledgment}

The author thanks Groq for providing free-tier API access that enabled the open-weight model experiments, and Google Cloud for Vertex AI credits used for the Gemini experiments. All experimental code and results are publicly available at \url{https://github.com/Johin2/iterative-code-repair}. The author declares no conflicts of interest.

\bibliographystyle{IEEEtran}
\bibliography{references}

\end{document}